\documentclass[twocolumn]{aa}
\usepackage{graphicx}
\usepackage{txfonts}
\begin{document}
   \title{2-D spectroscopy and modeling of the biconical
          ionized gas in NGC 4388\thanks{Based on observations 
          at SAO 6-m telescope (Russia)}}

   \author{S. Ciroi\inst{1}
          \and
          M. Contini\inst{1,2}
          \and
          P. Rafanelli\inst{1}
          \and
          G. M. Richter\inst{3}
          }

   \institute{Dipartimento di Astronomia, Vicolo dell'Osservatorio 2,
              I-35122 Padova, Italy\\
              \email{ciroi@pd.astro.it, piraf@pd.astro.it}
         \and
             School of Physics and Astronomy, Tel-Aviv University,
             Ramat-Aviv, Tel-Aviv, 69978 Israel\\
             \email{contini@post.tau.ac.il}
         \and
             Astrophysikalisches Institut Potsdam, An der Sternwarte 16,
             D-14482 Potsdam, Germany\\
             \email{gmrichter@aip.de}
             }

   \date{Received 9 September 2002; accepted 23 December 2002}


\def\ea{\rm et al.~}
\def\kms{$\mathrm{km\, s^{-1}}$}
\def\cm3{$\mathrm{cm^{-3}}$}
\def\Vs{$\mathrm{V_{s}}$}
\def\n0{$\mathrm{n_{0}}$}
\def\B0{$\mathrm{B_{0}}$}
\def\Te{$\mathrm{T_{e}}$}
\def\Fh{$\mathrm{F_H}$}
\def\erg{$\mathrm{erg~cm^{-2}~s^{-1}}$}
\def\Hb{H$\beta$~}
\def\Ha{H$\alpha$~}
\def\onel{$\lambda$}
\def\twol{$\lambda\lambda$}


\abstract{
We present recent results from spectroscopic data and modeling of the
biconical ionized gas in the Seyfert-2 galaxy NGC 4388.
A field of $\sim 2.6\times2.4$ kpc centered on the nucleus has been observed
by means of the modern technique of integral field spectroscopy.
The analysis of more than two hundred spectra allowed to study the physical
characteristics of the gas in the surroundings of the active nucleus.
The South-West ionization cone, revealed by the [\ion{O}{III}] \onel5007/\Hb
excitation map, shows high emission line ratios not completely supported by 
simple photoionization.
Composite models which account for the combined effects of photoionization and 
shock show that such high [\ion{O}{III}]/\Hb line ratios are emitted by low 
density (\n0=30 \cm3) gas inside large (D$>$1 pc) shocked clouds 
(\Vs=100 \kms) reached by a relatively low flux from the active nucleus.
The data of the VEELR in the North-East cone by Yoshida \ea (2002)
have been modeled.
The results confirm that  photoionization is the prevailing mechanism,
but nontheless weak shocks are under way between colliding clouds with
small ($<$1 pc) sizes and low densities \n0$\le$100 \cm3, moving outward at
relatively low velocities (\Vs=100~\kms).

\keywords{Techniques: spectroscopic -- Galaxies: nuclei --
Galaxies: Seyfert -- Shock Waves -- Galaxies: individual: NGC 4388
               }
   }

   \maketitle
%

\section{Introduction}

The prevailing picture of an AGN structure, the so-called Unified Model,
provides for a central supermassive Black-Hole (BH), whose gravitational
potential energy is the ultimate source of the AGN luminosity. Matter falling
into the BH loses angular momentum through viscous and turbulent processes in
an accretion disk, resulting into emission of photons  in the ultraviolet,
soft-X and hard-X ray wavelength domains (\cite{up95}).\\
According to the Unified Model the central source is surrounded by a dense
obscuring thick torus which is able to restrict the emergent radiation from 
the nucleus to a bipolar cone with an opening angle determined by the torus 
geometry (\cite{pog89}).
Imaging in optical emission lines shows in some cases high ionization regions
with a conical and/or biconical morphology. These so-called {\it ionization
cones} are direct evidence that radiation escapes anisotropically
(\cite{mul94}). \\
In this frame we have carried out an observational campaign to study the
circumnuclear and extranuclear environment of a spectroscopically selected
sample of Seyfert galaxies (\cite{raf95}) by means of the integral field 
spectroscopy. This technique offers much more advantages than the
traditional long-slit spectroscopy, because it allows us to observe
extended regions of nearby galaxies providing simultaneously recorded
spectral information (\onel) along two spatial directions
($\alpha$ and $\delta$). \\
Here we present results from the observations of the Seyfert-2 galaxy 
NGC 4388, a nearby (z=0.0084) highly inclined spiral galaxy
($i \sim$ 75\degr), believed located in the core of the Virgo cluster.
Indeed observed distortions in the outermost optical isophotes are a 
clue of the possible interaction between this galaxy and the other members 
of the cluster.\\
The IUE spectra of NGC 4388 revealed the presence of an UV source occulted 
from direct view and producing an ionizing continuum with a steep spectral 
slope (\cite{fo86}; \cite{kin91}; \cite{kin93}).\\
Narrow-band \Ha+[\ion{N}{II}] and [\ion{O}{III}] \onel5007 images pointed
out the presence of emission line regions with highly ionized gas
distributed into two opposite cones: a S-W cone extended to 30\arcsec~ and a
N-E plume extended up to 50\arcsec~ above the galactic disk.
Spectroscopic analysis confirmed the photoionization by the central
power-law continuum as the principal excitation mechanism
(\cite{pog88}; \cite{pet93}).

\begin{figure*}
\vspace{13.2cm}
\caption{The position of the spectrograph over the galaxy.
The $\times$ symbol indicates the assumed position of the central engine.}
\label{f1}
\end{figure*}

Kinematic data obtained with Fabry-Perot Interferometer (\cite{vei99})
revealed non-rotational blueshifted velocities in the extraplanar gas
northeast of the nucleus, likely produced by a bipolar outflow.\\
Using deep narrow-band imaging obtained with the Suprime-Cam mounted at the
Subaru Telescope, Yoshida \ea (2002; hereafter YOS02) have recently found a 
very large emission line region (VEELR) extended up to $\sim$ 35 kpc northeast 
of the nucleus. It consists of
many gas clouds or filaments, part of which (within 12 kpc) are clearly ionized by
the nuclear radiation. They claim that the most plausible origin of this
ionized gas is tidal debris due a past interaction with a gas-rich dwarf
galaxy.\\
Moreover at radio wavelengths NGC 4388 shows a two sided, asymmetric structure 
with a flat spectrum central component, offset by $\Delta\alpha$ =1.2\arcsec~
and $\Delta\delta$= 2.6\arcsec~ from the
apparent optical nucleus, a diffuse blob to the North and an elongated feature
to the South with a compact blob in it (\cite{hs91}).
Several radio maps have shown that this feature is clearly associated with the
S-W ionization cone, suggesting the possibility of an interaction
between the radio jet and clouds of the ionized gas (\cite{fal98}).
Such interaction generates shock fronts, which cannot be neglected in the
interpretation of the spectra.

In this paper we investigate the physical conditions in the
extended biconical ionized gas of NGC 4388 through the modeling of the
optical emission line and continuum spectra observed in different regions.
Both the ionizing radiation from the active nucleus and the shock effects
are consistently accounted for.
In Sect. 2 the observations and modeling of the S-W cone are presented.
In Sect. 3 the  N-E cone is considered and the VEELR is modeled on the basis
of YOS02 data. The continuum SED is discussed
in Sect. 4. Concluding remarks follow in Sect. 5.

\section{The South-West cone}

\subsection{Observations and data reduction}

NGC 4388 was observed in March 1998 at the 6-m telescope of the Special
Astrophysical Observatory (Russia) with the MultiPupil Fiber Spectrograph
(MPFS), an integral field unit made by an array of $16\times15$
microlenses coupled with a close-packed bundle of optical fibers, which carry
the incoming signal to the spectrograph.
Each of the 240 elements was covering a region of 1\arcsec$\times$1\arcsec~
corresponding to a total field of view of about 2.6$\times$2.4 kpc
(assuming H$_0$ = 75 km s$^{-1}$ Mpc$^{-1}$).
Two 600 sec exposures were taken using a TK-1024 CCD, which has a pixel size
of 24 $\mu$m, and orienting the largest side of the array at
P.A. $\sim$34\degr~ (Fig. \ref{f1}).
The grating was chosen to give a spectral range of $\sim$ 4400--7100 \AA,
with a resolution of 10 \AA\ and a dispersion of 5.3 \AA/px.

\begin{figure}
\resizebox{\hsize}{!}{\includegraphics{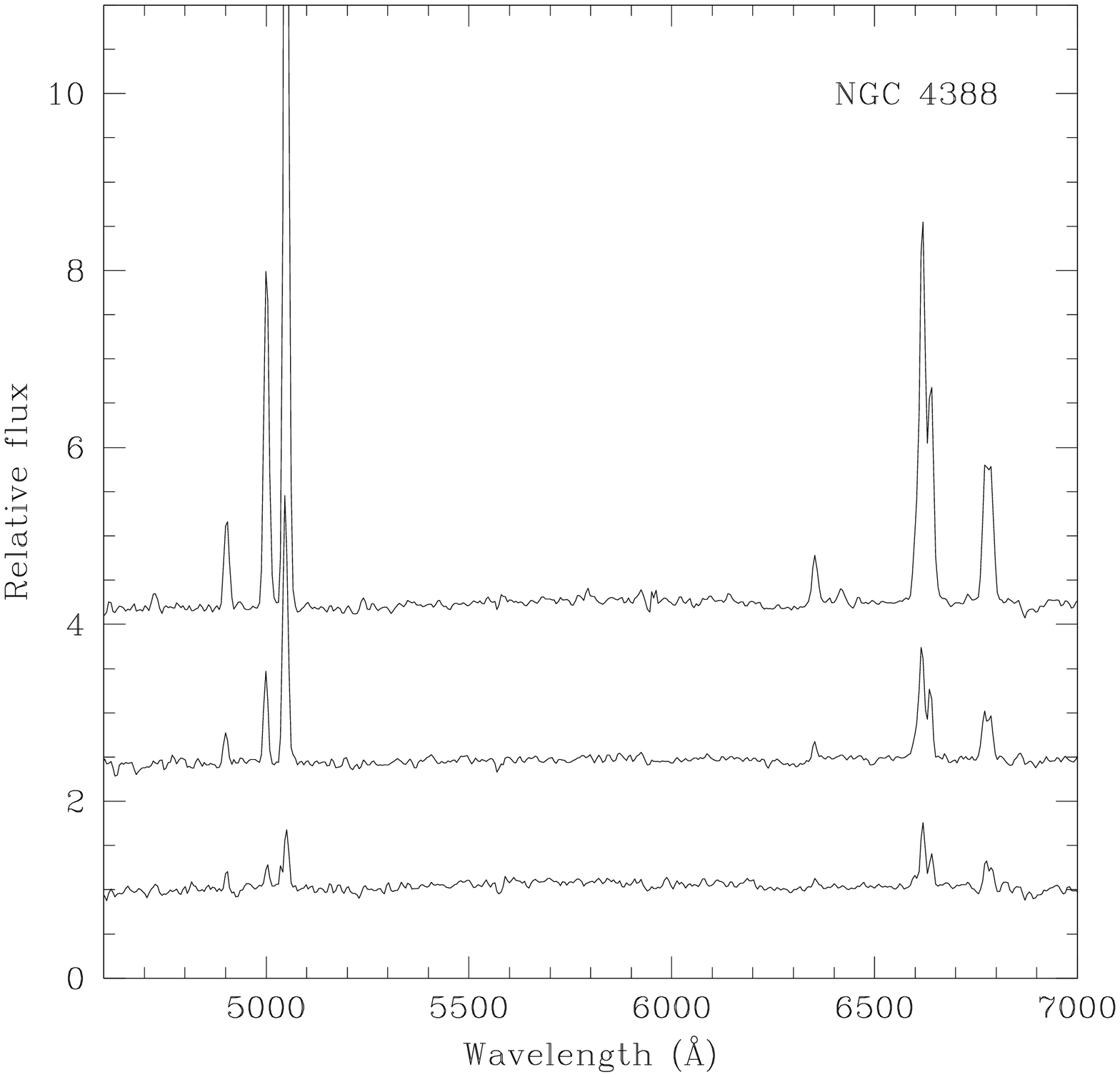}}
\caption{Examples of three 1-D spectra after the reduction.
They are representative of
different signal-to-noise ratio in the field of view.}
\label{f2}

\resizebox{\hsize}{!}{\includegraphics{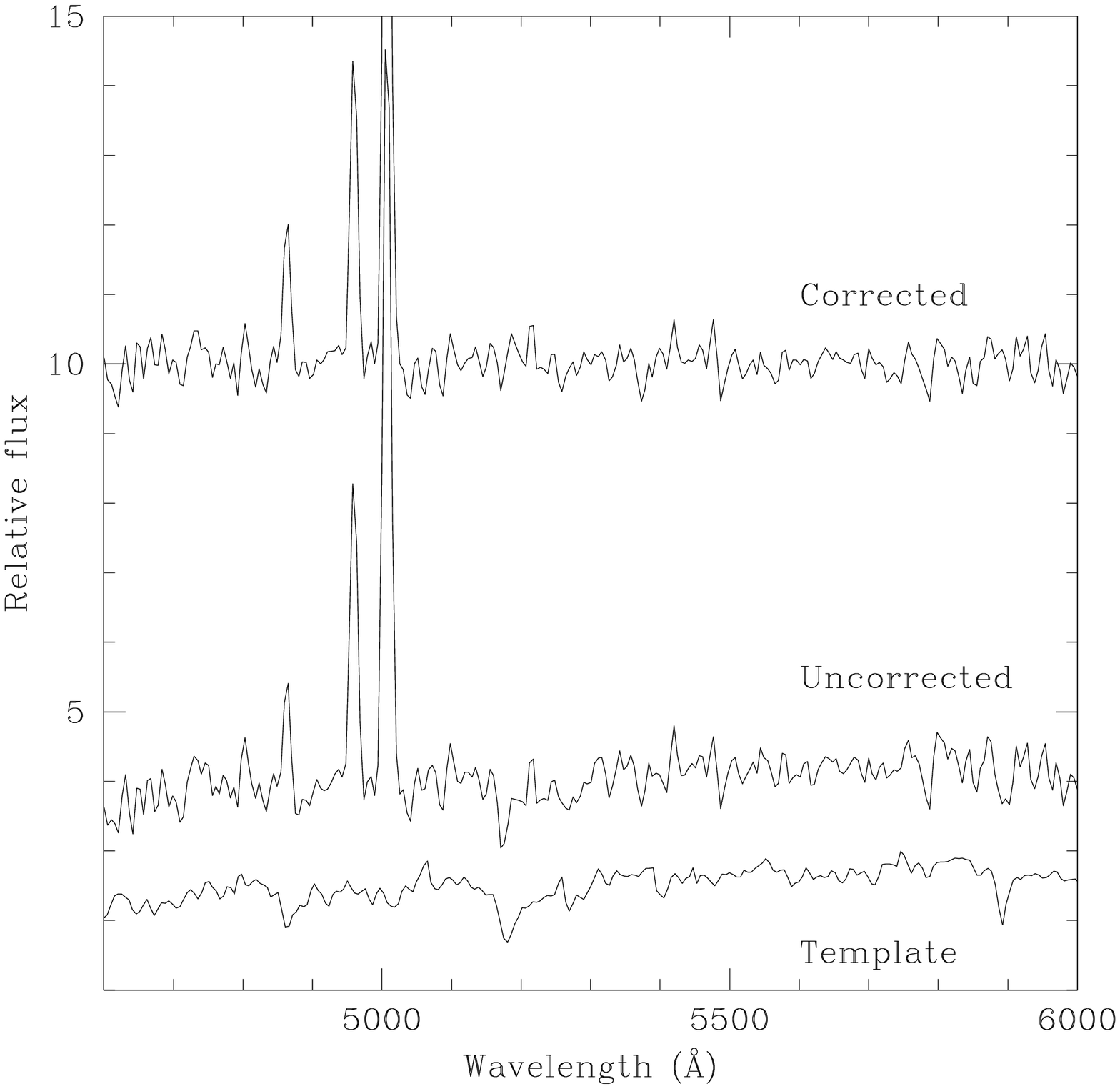}}
\caption{Example of the underlying starlight correction. An absorption-line
"template" spectrum, conveniently rescaled, is subtracted from one of the
240 observed spectra in NGC 4388. The result shows that metallic lines are
canceled out and the underlying \Hb absorption is removed.}
\label{f3}
\end{figure}

The reduction of the data was carried out by means of a software
package, developed within our group, and running under IRAF\footnote{
IRAF is distributed by the National Optical
Astronomy Observatories, which are operated by the Association of Universities
for Research in Astronomy, Inc., under cooperative agreement with the
National Science Foundation} environment.\\
After the usual steps of bias and dark subtraction, all the frames were
cross-correlated, in order to compensate for the shifts between them caused by
mechanical flexions of the instrument. Then a flat field was chosen as
reference image to obtain the geometrical scheme of the spectra arrangement:
the position, width and layout along the dispersion direction of every
spectrum stored in it were defined and the resulting scheme was subsequently
applied to all the images in order to extract for each of them the 240 1-D
spectra. The dispersion solution for wavelength calibration was determined
by fitting the line positions of He-Ne-Ar lamp exposures with low order
Chebyshev polynomial functions.
The sky subtraction was performed by averaging the output of 8 fibers devoted
to the night sky observation and placed at a
distance of 4.5\arcmin~from the center of the field of view.
Then the data were flux calibrated by observing the spectrophotometric
standard star HZ 44: all its flat-fielded, wavelength calibrated and sky
subtracted spectra were summed together in order to collect the total incoming
flux, to obtain the function necessary to convert counts in physical units and
to account for the sensitivity response of the system at different wavelengths.
Finally, cosmic rays were removed by combining corresponding spectra of
the two exposures of NGC 4388 and a correction for Galactic extinction
($\rm A_V = 0.08$) was applied. 
Fig. \ref{f2} shows some examples of the reduced
spectra chosen with different signal-to-noise ratio in the field of view.\\
The positions, widths and fluxes of all the emission lines detectable
3$\sigma$ over the continuum were measured using the IRAF task SPLOT.
In case of close blending like \Ha+[\ion{N}{II}] \twol6548,6583 or
[\ion{S}{II}] \twol6717,6731, better results in performing an interactive
multigaussian fit of the total profile could be obtained by using
the MIDAS package ALICE, whose input parameters (a first estimate of
amplitude, center and sigma of the fitting gaussian) are independently fixable.

\subsection{Data analysis}

In order to get accurate estimate of line fluxes and ratios it is necessary 
to take into account that emission lines in galaxies are generally affected, 
and in some cases even dominated, by the underlying starlight contribution.
It is possibile to remove the stellar features by subtracting an 
absorption-line "template" galaxy spectrum built by means of spectral 
synthesis techniques (see e.g. \cite{bon89}), or directly obtained from 
the observation of early-type spirals.
In our case the second approach was used, chosing a spectrum of a S0 galaxy 
extracted from our database which best matched the stellar features of most of 
NGC 4388 spectra, and following the suggestions given by \cite{ho93}. 
As a whole, the resulting correction was considered satisfactory (an example 
is given in Fig. \ref{f3}). 
The metallic lines (Mg I \onel5175, Fe I \onel5269 and Na I \onel5892) 
were almost always canceled out after the subtraction. Only in some cases
the stellar features of the template have been strengthened to better 
match the \Hb absorption, producing an overcorrection of the metallic 
absorptions. This can happen since stellar populations cannot be
omogeneus everywhere in the field of view.
The resulting variations of line intensities have been estimated 
weak ($< 10$\% for \Hb) in spectra with very bright emission lines, 
significant ($\sim 10-30$\%) in most of the spectra  and very important in 
few spectra (up to 60\%).

The emission line fluxes, measured after the starlight contamination
removal, were corrected for internal reddening by 
means of the theoretical Balmer decrement value \Ha/\Hb = 2.86 for Case B
recombination at electronic temperature \Te=10$^4$ K and assuming the
reddening law by \cite{mm72}.

Several images of the galaxy within the field of view were reconstructed 
at different wavelengths, as for instance the map of the visible 
continuum at 5500 \AA. The position of its maximum intensity is located to the 
N-W of the observed peak of the emission line maps, 
which we have decided to assume as position of the photoionization source 
(hereafter nucleus). We do not confirm the large offset between the location 
of the radio source with flat spectrum and the visible continuum peak 
indicated in Hummel \& Saikia (1991) and references therein: the radio source, 
the visible continuum and the emission lines have their maxima intensities 
close to each other (separation $\sim 1$\arcsec).
We have discarded the idea to adopt the position of the radio source, which 
should indicate the real location of the AGN, as nucleus in our field of view 
since the spatial resolution of our data is only 1 \arcsec/px and an accurate 
astrometric calibration of the integral field images is very hard to obtain.

\begin{figure}[b]
\vspace{8cm}
\caption{The ionization cone obtained by the reconstructed map of the
[\ion{O}{III}]/\Hb emission line ratios. The $\times$ symbol indicates the
position of the assumed nucleus (as in Fig. 1), OC is the maximum intensity
of the optical continuum at 5500 \AA, and RS is the approximate location of
the radio source with flat spectrum.}
\label{f4}
\end{figure}

The map of the [\ion{O}{III}] \onel5007/\Hb emission line ratios revealed
part of the S-W ionization cone (Fig. \ref{f4}).
The cone has the axis at P.A. $\sim$200\degr~ and an aperture angle of
$\sim$70\degr~, less than the value measured by Pogge (1988), probably
due to our smaller field of view.

\begin{figure}
\resizebox{\hsize}{!}{\includegraphics{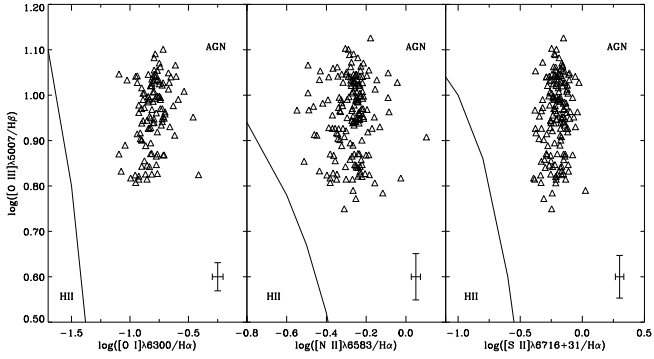}}
\caption{Diagnostic diagrams of the emission line regions inside the
ionization cone. The typical uncertainties of the ratios are indicated
by the error bars in the bottom-right corner of each plot.}
\label{f5}
\end{figure}

The gaseous regions within the ionization cone have been investigated.
Their intrinsic FWHM, estimated from the gaussian analysis of the emission 
lines after having removed the instrumental component ($\sim$10 \AA), range 
in the interval $\sim$200--700 \kms, with a median value of $\sim$480 \kms, 
velocities usually observed in the Narrow Line Regions of Seyfert galaxies.
The diagnostic emission line ratios [\ion{N}{II}] \onel6583/\Ha,
[\ion{O}{I}] \onel6300/\Ha, [\ion{S}{II}] \twol6716+6731/\Ha and
[\ion{O}{III}] \onel5007/\Hb (\cite{vo87}), show the typical properties of gas 
ionized by a non-thermal source (Fig.~\ref{f5}).

Moreover the spatial distribution of the ionization degree within the cone,
evaluated by means of the [\ion{O}{III}] \onel5007/\Hb
ratio as a function of the distance from the nucleus, indicates that
the gas is highly ionized even at large distances from the AGN. \\
In order to understand if the AGN could account for such observed ionization,
the \Ha luminosity of each spectrum was calculated from the reddening
corrected \Ha fluxes.
Then the number of ionizing photons needed to produce such luminosities,
$\rm Q_{ion}=7.3 \times 10^{11}~ L(H\alpha)$ photons sec$^{-1}$
(\cite{kenn98}),
was evaluated and later compared with $\rm Q_{nuc}^\star = Q_{nuc}~\Omega/4\pi$, 
where $\rm Q_{nuc}$ is the number of photons coming from the nucleus, diluted by 
the covering factor $\rm \Omega/4\pi$,  and $\Omega$
is the solid angle of a region seen from the nucleus.\\
By assuming that the nuclear source is an isotropic emitter of radiation
occulted along our line of sight, we attempted to estimate $\rm Q_{nuc}$ by
taking the average of the calculated number of photons ionizing the regions
surrounding the nucleus, after having removed the effect of the covering
factor ($\rm Q_{ion}~4\pi/\Omega$).
A value of $\rm Q_{nuc} \sim 4.87\times10^{52}$ photons sec$^{-1}$ was
obtained, in good agreement with the 4.98$\times$10$^{52}$ photons sec$^{-1}$
computed by \cite{col92} using the IUE spectrum of the galaxy and the flux
at 912 \AA.\\
The resulting $\rm Q_{ion}/Q_{nuc}^\star$ ratio concentrates around a value
of 2.5. This indicates that the active nucleus is the dominant ionizing source
for the regions within our field of view, in agreement with what observed
in the VO diagrams.
In spite of that, the ionization parameter $\rm U=Q_{nuc}/4\pi c r_0^2 N_H$,
calculated within the cone as a function of the distance
r$_0$ from the nucleus, ranges between $10^{-5.0}$ and $10^{-2.8}$, a result
clearly in contrast with the already pointed out high ionization level of
almost all the regions. A total hydrogen density
$\rm N_H=3\times10^2~cm^{-3}$ is assumed, on the basis of the the electron
density values estimated over the whole field, using the
[\ion{S}{II}] \twol6717/6731 ratio and assuming T=10$^4$ K (except
where the sulphur lines were not visible or extremely blended).
The problem of the missing data has been solved taking the average value of
independent linear interpolations of the [\ion{S}{II}] ratios along the two 
orthogonal directions of the spectral array. The resulting N$_e$ distribution 
presented values $<$10$^3$ cm$^{-3}$, with a median of $\sim$300 cm$^{-3}$. 
Of course due to the uncertainties of the [\ion{S}{II}] emission line strengths,
the estimation of the electron density is not very accurate and just a simple
decrease  by a factor of 10 can account for the ionization parameter in  the
regions far from the nucleus.

\subsection{The models}

The contribution of the shocks to the line intensities is tested by comparing
calculated  with observed spectra.
The code SUMA (\cite{cv01}),
which  accounts consistently for the photoionization flux from an
external source and for the shocks is adopted.

The general model assumes matter bound gaseous and dusty clouds, which move
outwards from the source and emit the line and continuum spectra.
A shock front forms on the outer edge of the clouds, while the inner edge is
reached by the radiation flux from the source.

The  input parameters are : the radiation flux intensity, \Fh~
(in photons $\mathrm{cm^{-2}~s^{-1}~eV^{-1}}$ at 1 Ryd), the spectral
index, $\alpha$ (= 1.5 for all models), the shock velocity, \Vs, the
preshock density, \n0, the preshock magnetic field, \B0
(= $10^{-4}$ gauss), the dust-to-gas ratio, d/g, and the geometrical
thickness, D, of the emitting clouds. Cosmic relative abundances (\cite{all73})
are adopted.

Composite models lead to a complex distribution of the temperature and the density
throughout the clouds. In radiation dominated (RD)   models the effect of the flux
from the active center (AC) prevails on the shock, while in shock dominated (SD)
models the flux is absent (\Fh = 0).
When shocks and photoionizing fluxes act on the opposite  sides of the cloud,
two regions of relative high temperature appear near the edges (see Sect. 2.4).
Higher temperatures   correspond to higher shock velocities (T $\propto$ \Vs$^2$),
while the radiation flux cannot heat the gas to $>$ 2 - 3 10$^4$ K.
This temperature leads  to  high O$^{+2}$/O  oxygen fractional abundance,
so  the  [OIII]  line  strength depends on \Fh.

The cooling rate ($\propto$ n$^2$, where n is the density of the gas)  increases
with compression downstream of the shock leading to  cool gas in the internal
region of the cloud.
The larger the cool region the strongest the low ionization level lines.
The relative sizes of the hot and cool gas regions depend, therefore, on
the shock velocity, on the intensity of the flux from the AC,  on the preshock
density, and on the geometrical thickness of the clouds.

A large grid of models  is calculated in order to select the most
suitable parameters to the observational evidences for each region in the galaxy.
Recall that a first choice  is roughly obtained considering that
 the [\ion{O}{III}] \onel5007 line  flux depends on the intensity of the nuclear
radiation flux, the [\ion{S}{II}] \twol6717/6731 line
ratio depends on the density, as seen before, and
the intensity ratios of high to low ionization
lines depend on the geometrical thickness of the clouds and on the shock
velocity (see Contini \& Viegas 2001).
The input parameters are then changed consistently in order to obtain the
best fit of all the observed data.

\subsection{[\ion{O}{III}]/\Hb and [\ion{O}{I}]/\Hb}

The observations cover the  SW cone  providing  line ratios
in many different regions.
We focus on [\ion{O}{III}] \twol4959+5007/\Hb and
[\ion{O}{I}] \twol6300+6363/\Hb line ratios, because [\ion{O}{III}] lines
are generally strong in Seyfert 2 galaxies and are emitted
by gas ionized by the radiation flux from the active nucleus, while [\ion{O}{I}]
lines are emitted from gas at lower temperatures and are generally strong in
the presence of shocks.\\
Moreover, \cite{vbc99} claim that dynamical processes such as entrainment by
AGN-powered radio jets generally have a stronger effect on the kinematics of
the highly ionized [\ion{O}{III}] emitting gas than on those of the low-ionization
\Ha emitting material.

\begin{figure}[b]
\resizebox{\hsize}{!}{\includegraphics{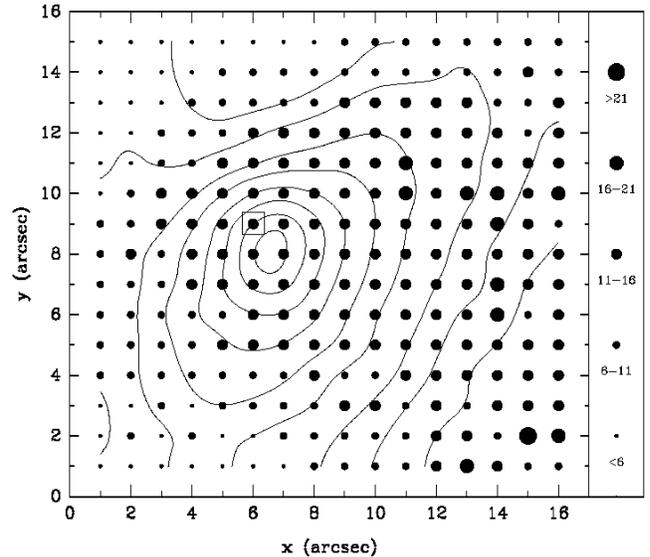}}
\caption{Contour map of the visible continuum at 5500 \AA. The isointensities
correspond to the values: 1.2, 1.5, 1.8, 2.2, 2.5, 2.8, 3.0, 3.15, 3.25
($\times$ 1.e-15 \erg).
Filled circles correspond to the values of the observed
[\ion{O}{III}]\twol 4959+5007/\Hb ratios.
The small "squared window" located at (6,9) indicates
the assumed position of the ionization source.
The map is orientated like in Fig. 4.}
\label{f6}
\end{figure}

Therefore, by modeling the observed spectra it is possible to determine the
distribution of the radiation flux intensity and of other parameters, e.g. the
shock velocity of the gas, the density, and the geometrical thickness of the
clouds in the observed region.
The reliability of the models is checked comparing the observed
spectral energy distribution (SED) of the continuum with model results
(see Sect. 4).

The observed [\ion{O}{III}]/\Hb line ratios overlaid to the continuum of the
galaxy at 5500 \AA\ are shown in Fig. 6 in a synthetic representation.
The [\ion{O}{III}]/\Hb distribution is not smooth.
Particularly,  high [\ion{O}{III}]/\Hb ratios appear in some regions far
from the active center. This suggest that the clouds in those regions have 
abnormal intrinsic conditions.

\begin{figure}
\resizebox{\hsize}{!}{\includegraphics{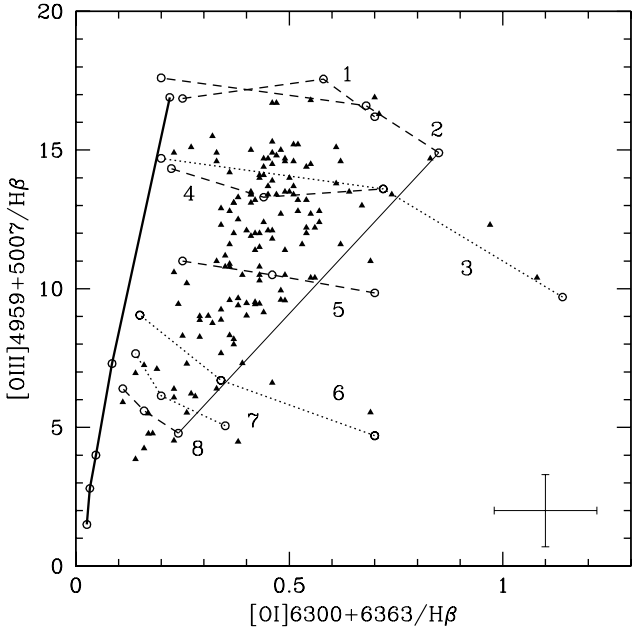}}
\caption{The comparison of model calculation with
observations (filled triangles) for [\ion{O}{III}]\twol 4959+5007/\Hb
vs [\ion{O}{I}]\twol 6300+6363/\Hb.
See text for a detailed description.}
\label{f7}
\end{figure}

In Fig. 7 [\ion{O}{III}]/\Hb versus [\ion{O}{I}]/\Hb from the observations
(black triangles) is compared with model calculations.
All the models correspond to \Vs=100 \kms.
Higher velocities are less fitting, and velocities lower by a factor $<$ 2
lead to similar results.
The thick solid line  refers to shock dominated models
 with line ratios  increasing  with decreasing D and \n0.
The bulk of the data, however, is scattered   throughout a region
of the diagram better fitted by  radiation dominated  models.
The thin solid line determines roughly the right  edge of this region.
The numbers in Fig. 7 refer to  Table~\ref{tbl1}, where the models are 
described.  
The dashed lines connect models calculated with different D 
(increasing from left to right)
but all corresponding to the same flux intensity \Fh, while the
dotted lines connect models calculated with different \Fh~
(increasing from left to right) and corresponding to the same D. 
Fig. 7 shows that the trend of both types of 
models  is different from  the observed one, i.e.
[\ion{O}{III}]/\Hb increasing with [\ion{O}{I}]/\Hb.

The increasing trend of the line ratios can be explained
either by a slightly larger D and stronger \Fh~(models 1 in Table 1)
or by very large  clouds (D $\geq$ 1 pc, models 2 in Table 1), 
corresponding to a lower  density,
and a flux intensity not exceeding the  characteristic values
calculated for most of the observed regions inside the cone.
The other  observed lines at each position are too few and too uncertain
to  constrain the models, therefore, we suggest that the
clouds   reached by a strong nuclear flux (model 1) are located in the central
region of the cone, while clouds showing the same high [OIII]/\Hb ratios,
but characterized by a low density and a large geometrical thickness (model 2),
are located in some  external regions of the cone.
Extended  clouds of rarefied gas appear in the  terminal region of jets
and winds.

\begin{table}
\centering
\caption{The models in Fig. 7. ~\label{tbl1}}
\begin{tabular}{lllllllll}\\ \hline
\   & \n0 (\cm3)   & D (10$^{17}$ cm)  & \Fh$^1$   \\ \hline
\ 1  & 50.         & 9, 10, 12                & 1.5 \\
\ 2   & 30.      & 30, 40, 50               & 0.9\\
\  3  & 50.       & 7.                & 0.7, 1.0, 1.2 \\
\  4  & 50.       & 5, 6, 7                & 1. \\
\  5  & 50.       & 3.5, 4, 5               & 0.7\\
\  6  & 100.      & 0.6               & 0.8, 1.0, 1.2 \\
\  7  & 100.      & 0.4               & 0.6, 0.8, 0.9 \\
\   8  & 100.       & 0.2, 0.25, 0.3            & 0.6 \\
\hline\\
\end{tabular}

$^1$ in units of 10$^9$ photons cm$^{-2}$ s$^{-1}$ eV$^{-1}$ at 1 Ryd

\end{table}

The distribution of \Fh~ in the different regions of the cone, as it
results from modeling is shown in Fig. 8.
The range of the corresponding ionization parameters U is
2 10$^{-5}$ - 5 10$^{-4}$.

\begin{figure}
\resizebox{\hsize}{!}{\includegraphics{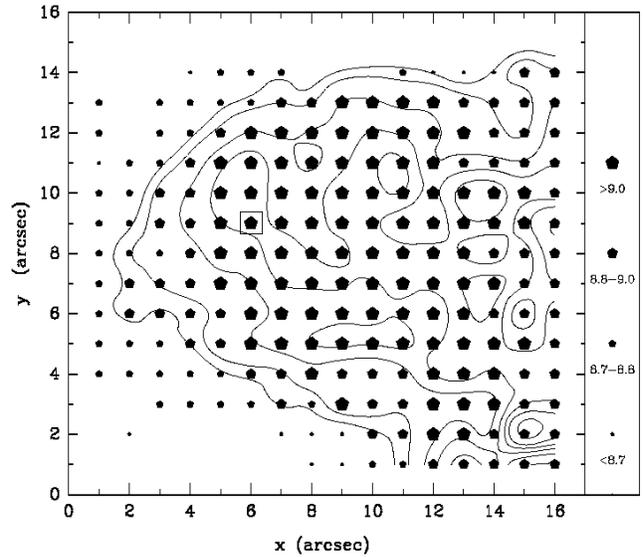}}
\caption{The grid of the $\rm F_H$ values (filled pentagons)
overlaid to the contour map of the ionization cone. The isointensities
correspond to the [\ion{O}{III}]\twol 4959+5007/\Hb ratios:
9.5, 10.5, 12.5, 14, 15.5, 17, 19.
The map is oriented as in Fig. 4. The square is like in Fig. 6.}
\label{f8}
\end{figure}

To better understand the results obtained above,
the distributions of the electron temperature (dotted line)
and density (solid line) throughout a cloud are given for different models
in Figs. 9  and 10.
Each figure is divided by a solid vertical line in two halves. The horizontal
axis scale is logarithmic and symmetric to have a comparable view of the two
sides of the cloud. The shock front is on the left and the radiation flux
from the source reaches the right edge.

\begin{figure}[h]
\resizebox{\hsize}{!}{\includegraphics{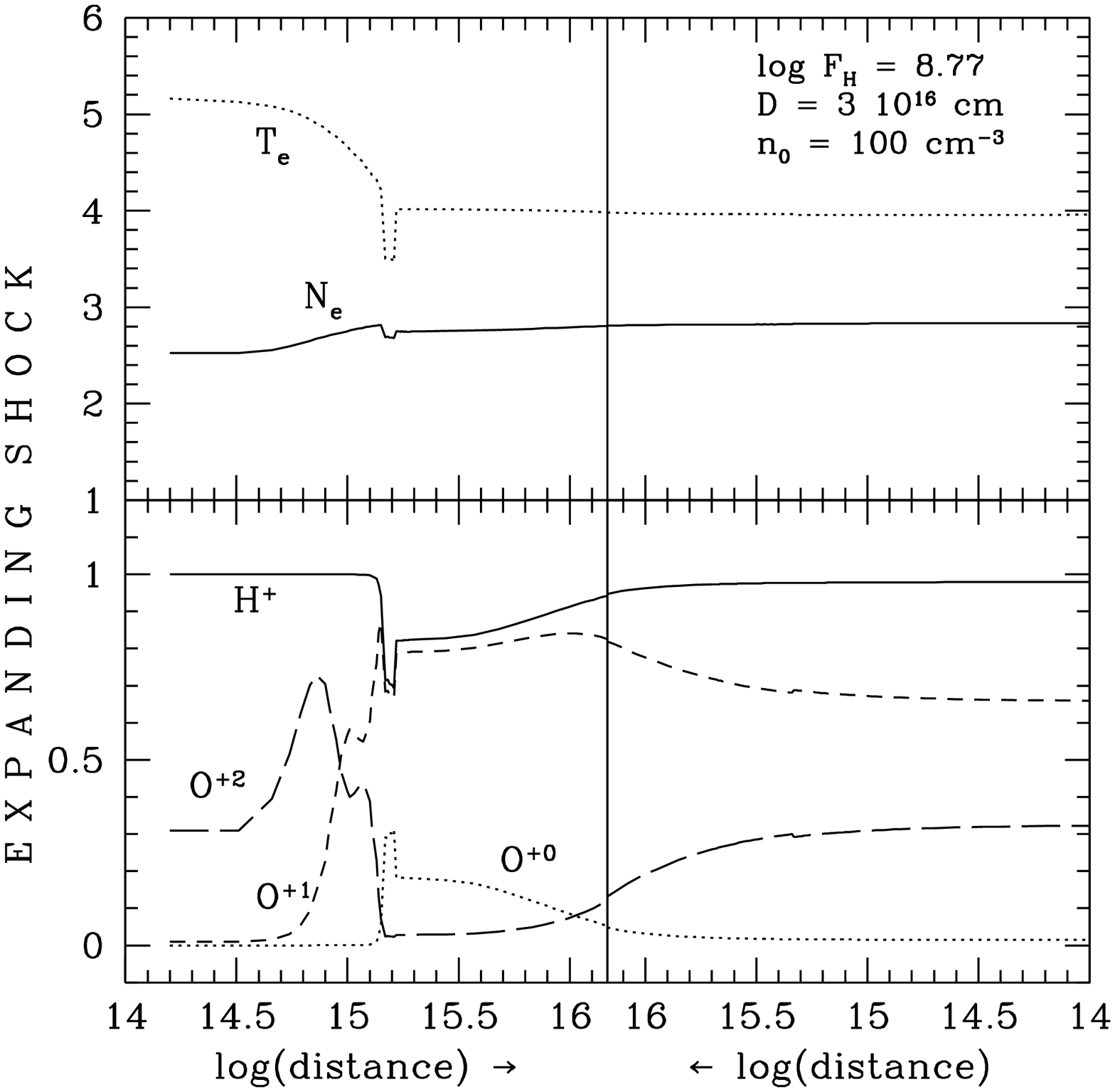}}
\caption{The case of a small D and relatively low $\rm F_H$.
The distribution of the temperatures and densities (upper panels)
and  of the most significant ion fractional abundances (lower pannels)
throughout the clouds for different models (see text).}
\label{f9}
\resizebox{\hsize}{!}{\includegraphics{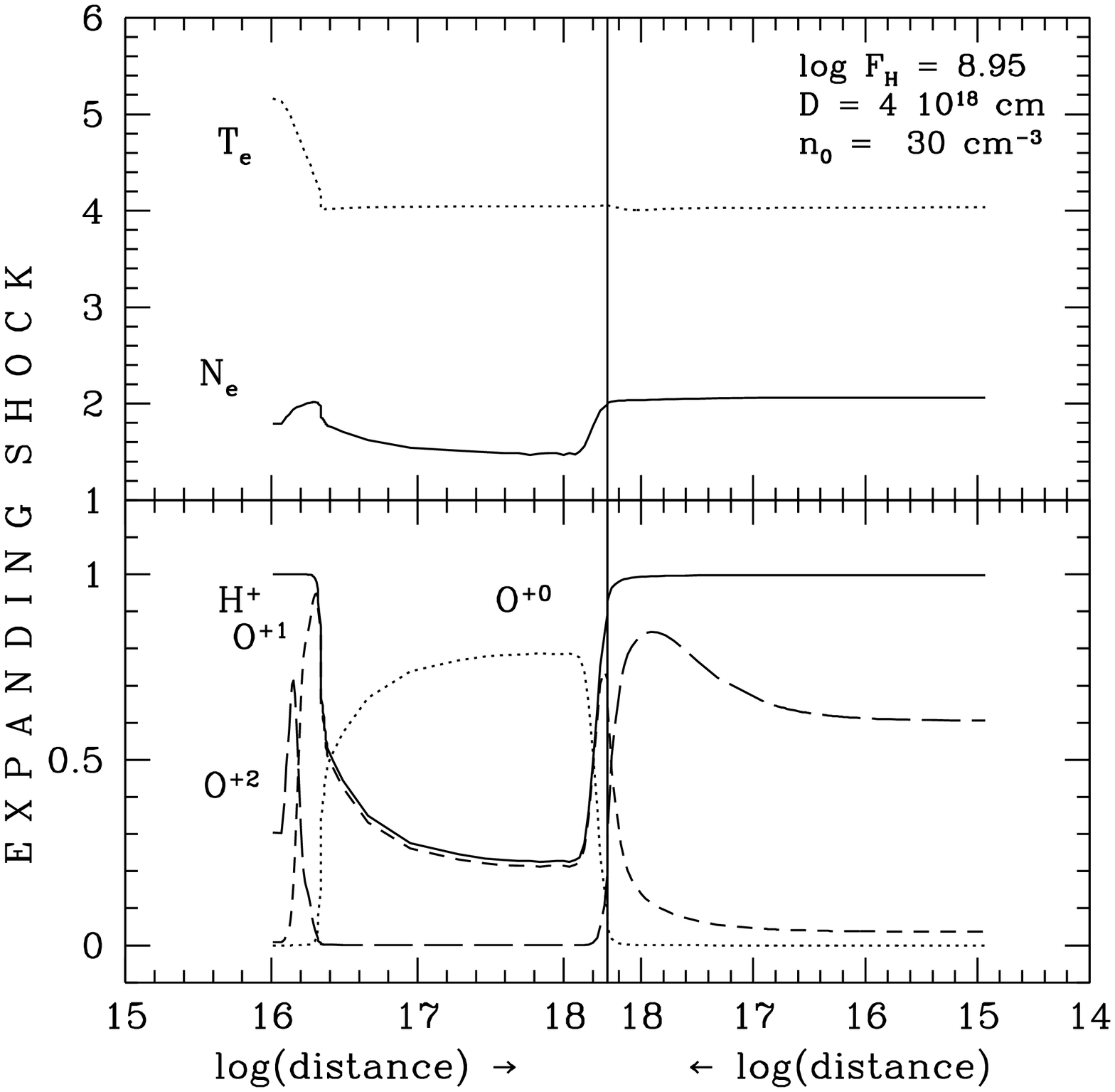}}
\caption{The same as for Fig.~\ref{f9} in the case of a large D and low n$_0$.}
\label{f10}
\end{figure}

The geometrical thickness D is a crucial parameter to low-level and,
particularly, to neutral lines, because they are emitted from rather cold gas.
Therefore, in the bottom pannels the distribution of the fractional abundance
of $\rm O^{+0}$/O (dotted line), $\rm O^{+1}$/O (short-dash line),
$\rm O^{+2}$/O (long-dash line), and $\rm H^{+1}$/H (solid line) have been
plotted.  
Fig. 9 corresponds to D= 3 $10^{16}$ cm  (0.01 pc) and Fig. 10 to
D= 4 $10^{18}$ cm (1.3 pc).
It can be  noticed that two zones of gas emit the [\ion{O}{III}] lines, one from
the side of the shock, where collisional ionization prevails, and the other from
gas ionized by radiation.
The ionization rate due to the photoionization flux from the nuclear source,
dominates in a large region of the cloud, particularly in models calculated
with \Vs=100 \kms, where the gas density and optical thickness are rather low.
A large region of cool gas appears inside clouds corresponding to a large
geometrical thickness and low density, leading to relatively strong [\ion{O}{I}]
lines.

\subsection{[\ion{S}{II}]/\Hb}

Fig.  11 shows [\ion{S}{II}] \twol6717+6731/\Hb versus
[\ion{S}{II}] \twol6717/6731.
The low ionization lines are strongly affected by shocks.
For models accounting for the shock a large range of electron densities
follows from compression in the downstream region.
The lines, calculated integrating through the geometrical thickness of the
clouds, account for the different conditions of the gas (see section 2.4).

\begin{figure}[b]
\resizebox{\hsize}{!}{\includegraphics{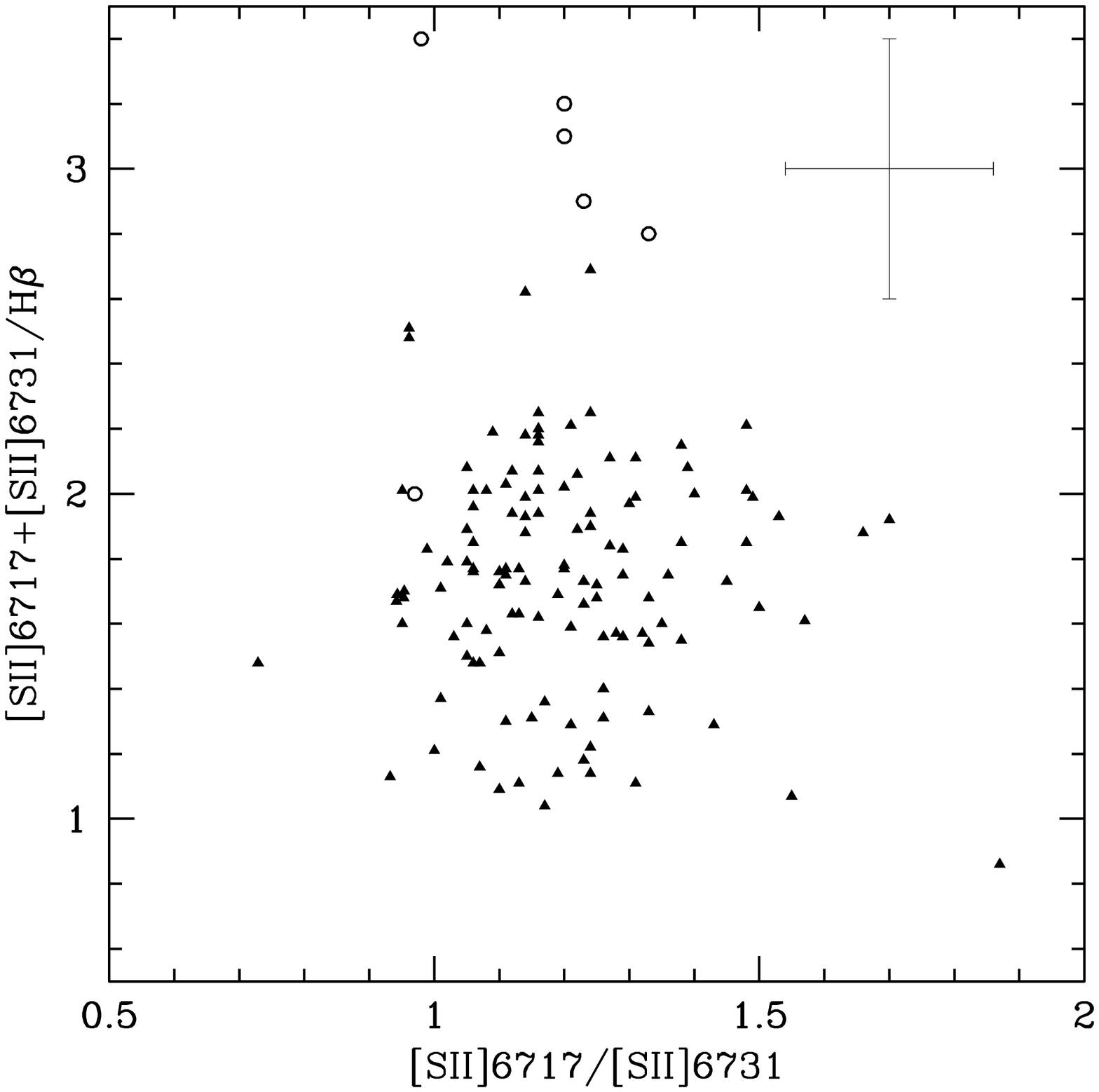}}
\caption{The same as for  Fig.~\ref{f7} for [\ion{S}{II}]\twol 6716+6731/\Hb vs
[\ion{S}{II}]\twol 6716/6731.}
\label{f11}
\end{figure}

Some significant models, which are the same as those adopted to explain the 
[\ion{O}{III}]/\Hb ratios in  Fig. 7, are indicated by open circles. 
They show a good agreement to the observed [\ion{S}{II}]
\twol6717/6731 line ratio between 1 and 1.3 in Fig. 11.
The low density model
gives [\ion{S}{II}] \twol6717/6731 $\leq$ 1.4.
Notice, however, that the [\ion{S}{II}] \twol6717+6731/\Hb ratios are
higher by a factor $<$ 2 than the data for most models, indicating that the
S/H relative abundance is lower than cosmic.
This is a common characteristic to Seyfert 2
galaxies where sulphur is depleted from the gaseous phase.
The models are calculated adopting d/g (by number) =  10$^{-14}$ - 10$^{-13}$,
indicating that dust is indeed present in the NGC 4388 cone region.

Reminding that compression increases with \Vs, models calculated with
higher densities and velocities (\Vs=300 \kms) fit the data
corresponding to low [\ion{S}{II}] \twol6717/6731 ($<$1), although their
observed errors are large.
Finally, models with \Vs=500 \kms~ hardly fit any data,
suggesting that high velocity clouds are relatively few (see Sect. 4).

The results show that the bulk of the shock velocities  is lower than the
velocities corresponding to the  FWHM of the emission lines (200-700 \kms),
 indicating that strong shocks do
not form in the observed region of NGC 4388, but rather low shock velocities
\Vs=100 \kms~ and low densities \n0$\leq$100 \cm3 dominate.

\section{The North-East cone}

We now consider the N-E cone and the very extended emission-line region
(VEELR) discovered by YOS02
in order to have a larger view of the ionization conditions in NGC 4388.
The nuclear region of the galaxy has been successfully investigated by
Petitjean \& Durret (1993), but only a few emission lines were observed at
21\arcsec, 25\arcsec, and 38\arcsec~ from the nucleus.

YOS02 observed the [\ion{O}{III}] \onel5007 and
\Ha+[\ion{N}{II}] \twol6548,6583 fluxes from
different clouds in the VEELR, which extends  northeastward of the galaxy,
very far outside the galaxy disk.
The  ionization cone is traced by the N-E plume and a high
ionization cloud located at 10\arcsec-15\arcsec~ northwest of the nucleus.
They claim that there is a sudden change in the excitation
state of the gas around the line
at P.A. 65\degr, which coincides with the extrapolation of the northern edge
line of the S-W cone.
Notice that these authors assume as distance of NGC 4388, the distance of the
Virgo cluster (16.7 Mpc), which is different from that one used by us
for the S-W cone (33.6 Mpc).
Therefore, all the distances should be scaled by a factor of 2.

In Figs. 12, 13, and 14 we compare the results of models calculated
with the code SUMA with the YOS02 data.
The errorbars refer to the typical uncertainty of 20 \% estimated
by YOS02.
In all the diagrams the fluxes of [\ion{O}{III}] versus \Ha+[\ion{N}{II}]
(in \erg) are presented.
Recall that models are calculated at the emitting nebula, while data are
observed at Earth. Therefore, we have multiplied the data by d$^2$/r$^2$,
where $r$ is the distance of each cloud from the center of the galaxy and $d$
is the distance of NGC 4388 from Earth.

All models are calculated adopting a geometrical thickness, D = 3 pc, and
\B0 = 10$^{-5}$ gauss, which is  similar to the magnetic field in the ISM
(\B0 = 10$^{-6}$-10$^{-5}$ gauss).

\begin{figure}
\resizebox{\hsize}{!}{\includegraphics{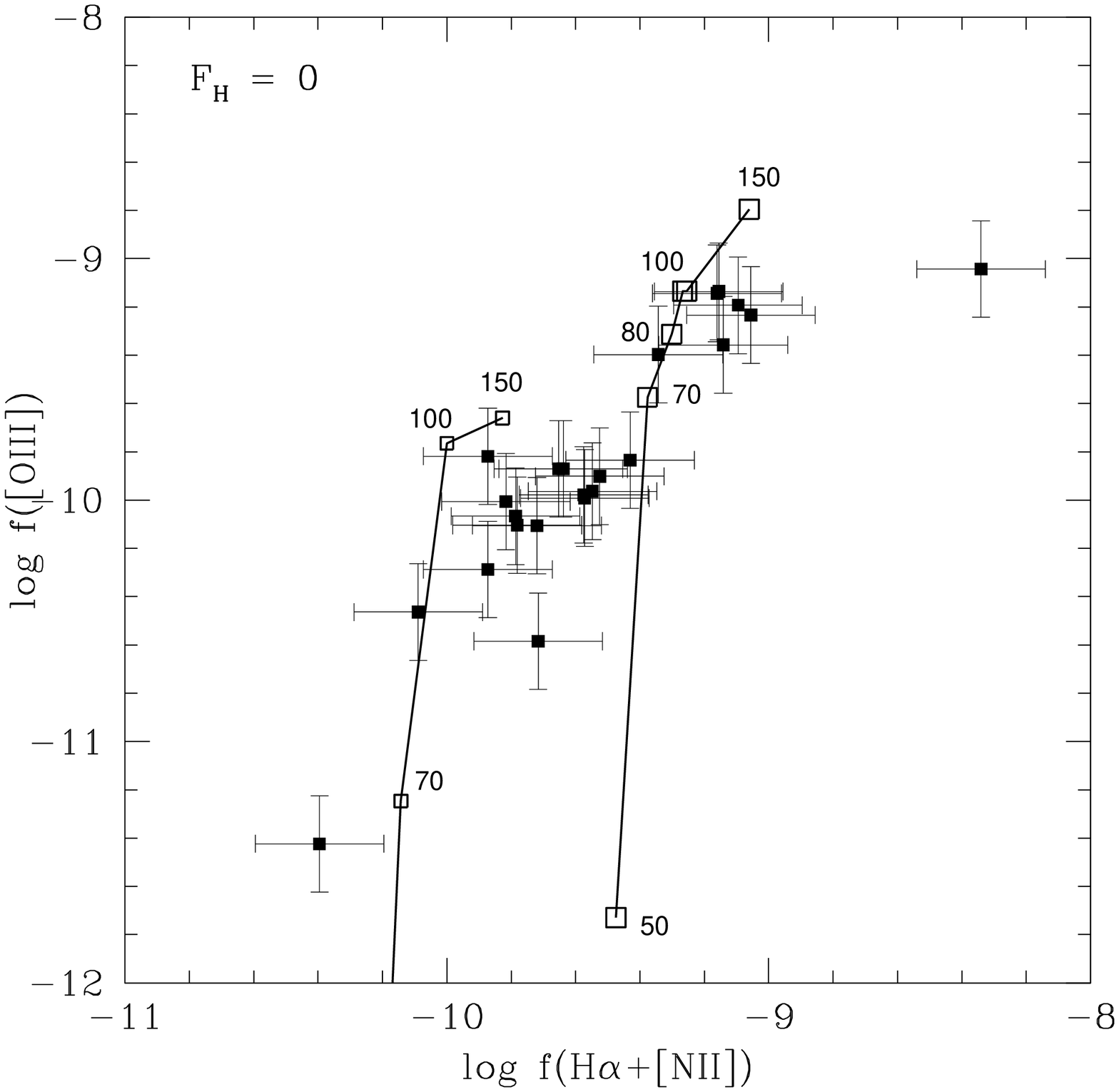}}
\caption{The comparison of model results (corresponding
to \Fh=0.)
with the observations for [\ion{O}{III}] \onel5007 (\erg) versus
[\ion{N}{II}]+\Ha (\erg) in the VEELR. Symbols are given in the text.}
\label{f12}

\resizebox{\hsize}{!}{\includegraphics{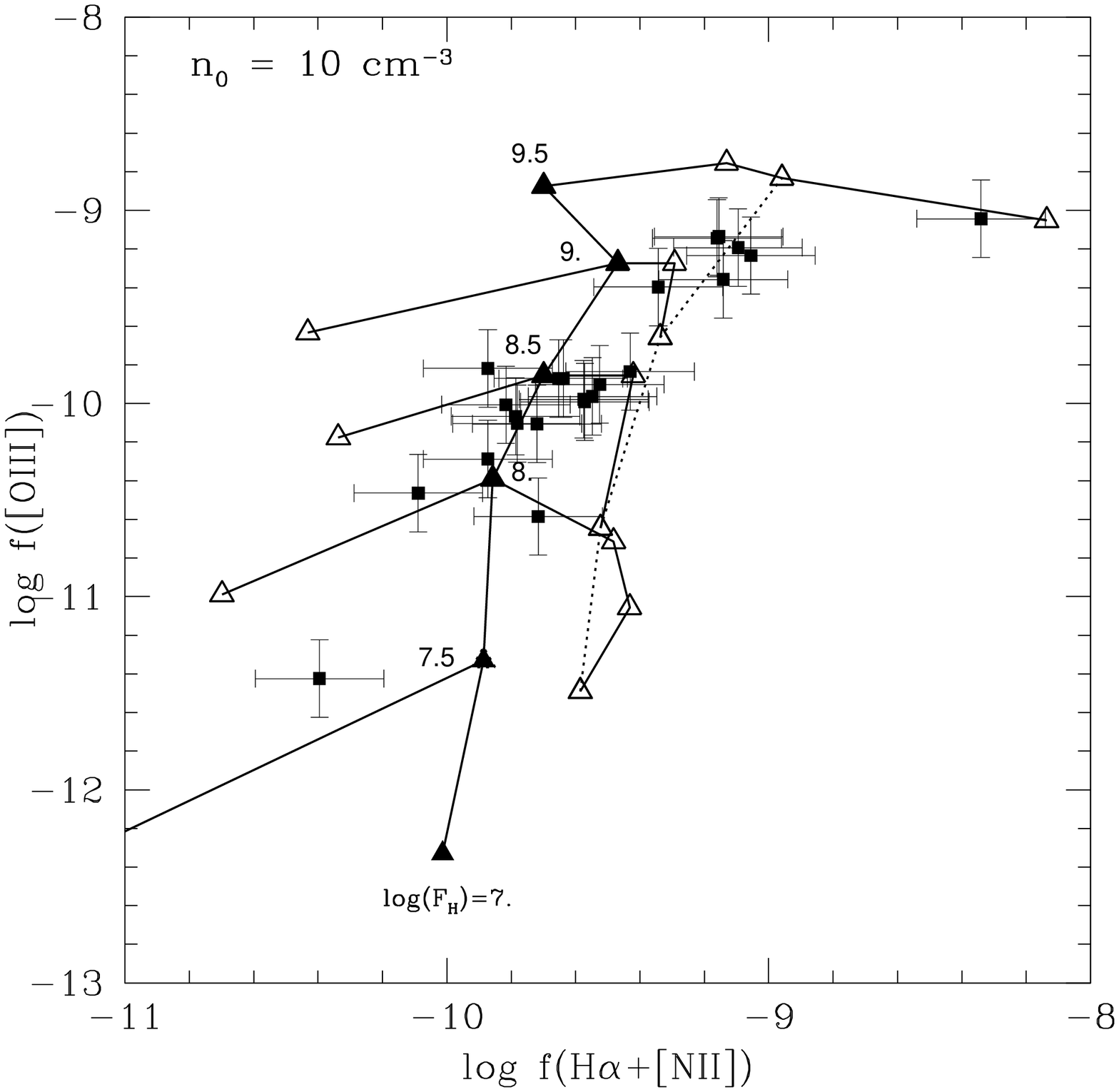}}
\caption{The same as for Fig.~\ref{f12} for composite models
calculated with \n0 = 10 \cm3.}
\label{f13}
\end{figure}

\begin{figure}
\resizebox{\hsize}{!}{\includegraphics{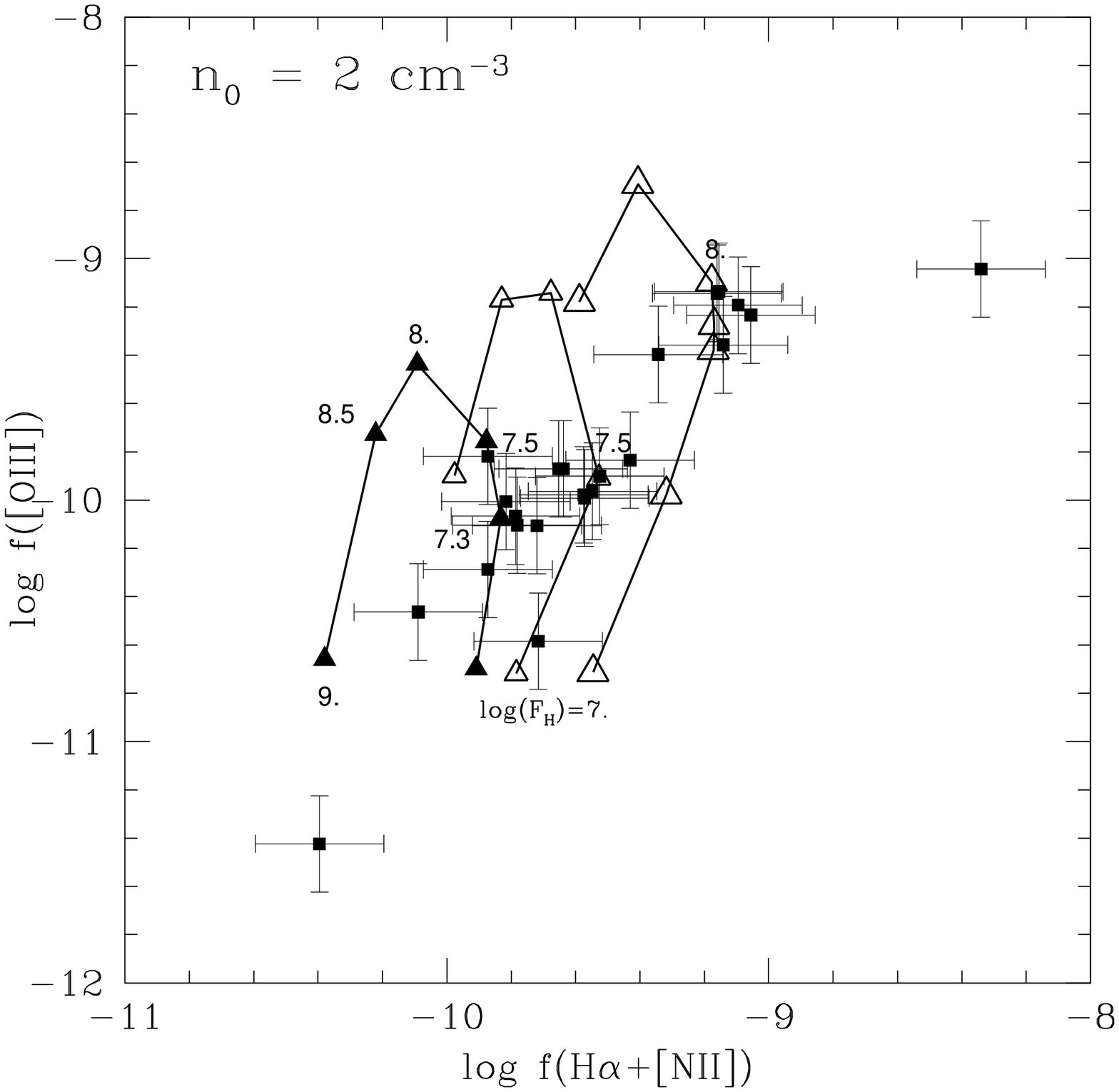}}
\caption{The same as for Fig.~\ref{f13} for composite models
calculated with \n0 = 2 \cm3.}
\label{f14}

\vspace{21pt}

\resizebox{\hsize}{!}{\includegraphics{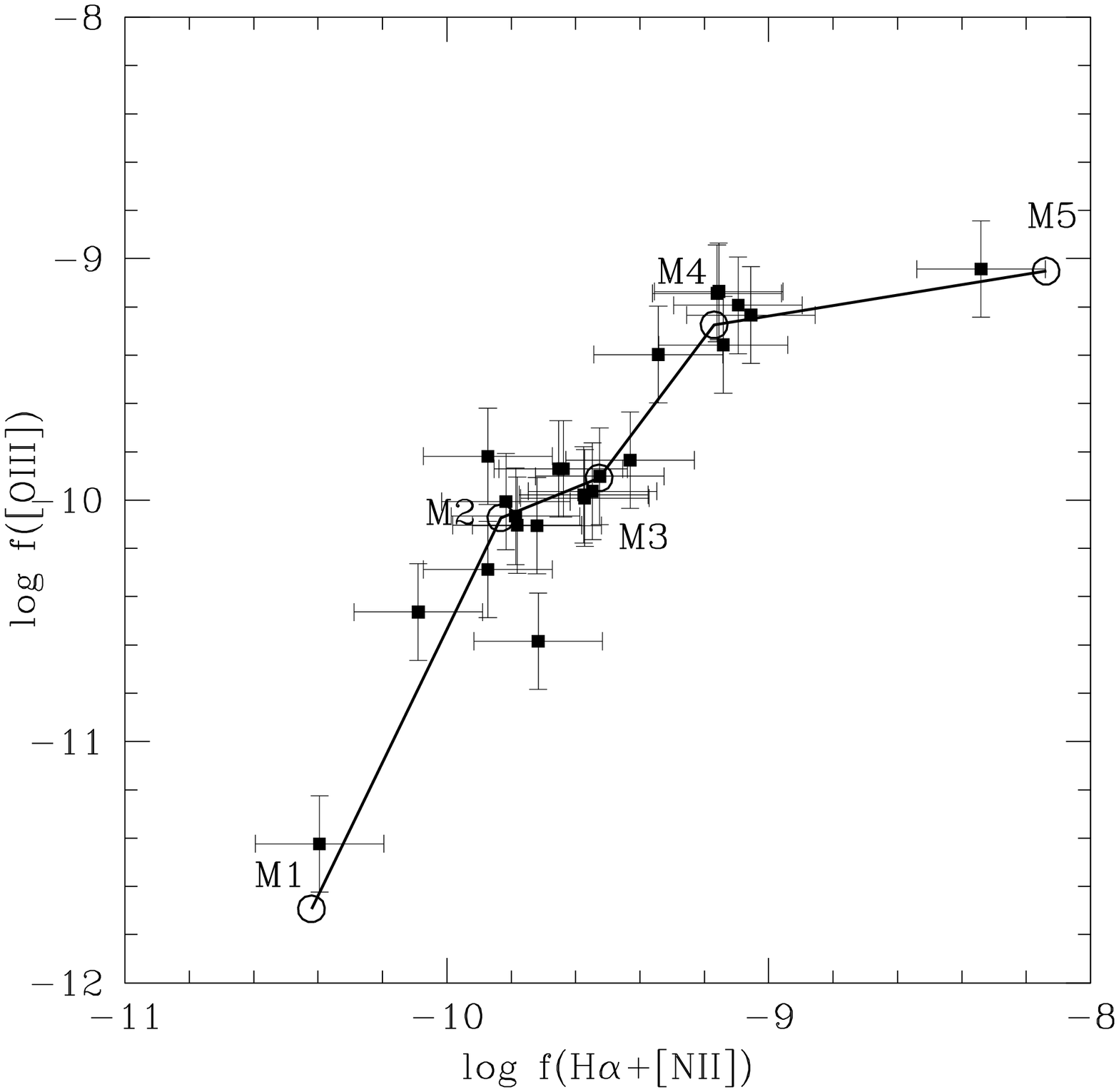}}
\caption{The comparison of selected model results
with the observations for [\ion{O}{III}] \onel5007 (\erg) versus
[\ion{N}{II}]+\Ha (\erg) in the VEELR (see text).}
\label{f15}
\end{figure}

Shock dominated models, calculated considering only  heating and ionization
of the  clouds by the shock (\Fh=0), are presented in Fig. 12.
The models  are labeled by the shock velocity in \kms.
Large squares correspond to \n0=10 \cm3, small squares to 
\n0=2 \cm3. Relatively  low preshock densities lead to the best agreement
with the data, however,
the SD models give a poor fit to the observed trend.
This confirms  YOS02 suggestion  that gas excitation depends also 
on the flux from the active nucleus even at such large distances.
We have used, therefore,  RD models to fit the data.
 Recall that RD models are also composed because they account for  the shock,
even if the effect of the photoionizing flux prevails.

We present in Figs. 13 and 14  radiation dominated models calculated with
different velocities and
fluxes,  adopting  \n0 = 10 \cm3 and 2 \cm3, respectively.
Higher densities do not fit.
In Fig. 13 black triangles indicate models calculated with \Vs=50 \kms,
white triangles in the left side  of the diagram corresponding
to \Vs=30 \kms and at the right  to \Vs=70-90 \kms.
The models are labeled by the values of log(\Fh).
The maximum [\ion{N}{II}]+\Ha value is explained by \Vs=150 \kms.
Dotted lines join the results of models calculated with \Vs = 90 \kms.
Notice that, roughly, all the data can be reached by the models
represented by the grid.

In Fig. 14 the models are calculated with \Vs=50 \kms~ (black triangles),
\Vs=70 \kms~ (small white triangles), and \Vs=100 \kms~ (large white triangles).
The adopted fluxes  range between log \Fh = 7. and  9. It can be noticed that
the   calculated trends do not correspond to the observed one, however,
some data are well reproduced by the  models. In particular, the maximum and
minimum observed  [\ion{N}{II}]+\Ha are not explained by these models.

We have  chosen  the best fitting models from Figs. 13 and 14  and
we present in Fig. 15  consistent modeling of the  entire dataset.
The selected models represented by open circles are  described in
Table~\ref{tbl2}.

\begin{table}
\centering
\caption{The best fitting models.\label{tbl2}}
\begin{tabular}{lllllll}
\hline\hline
               & M1 & M2 & M3 & M4 & M5 \\
\hline
\   \n0 (\cm3) & 2. & 2.  & 2.  & 2. & 10. \\
\  \Vs  (\kms) & 35. &50.  & 70. & 100. & 150. \\
\  log(\Fh)    & 6.5  & 7.3  & 7.5  & 7.9  & 9.5 \\
\hline
\end{tabular}
\end{table}

The fluxes of [\ion{O}{III}] and \Ha+[\ion{N}{II}] are given in Fig. 16 as
function of distance from the center. The data refer to the fluxes observed at
the clouds. Fig. 16 shows that the line fluxes decrease with distance, however,
different conditions coexist in the VEELR  leading to the large scattering.

\begin{figure}[b]
\resizebox{\hsize}{!}{\includegraphics{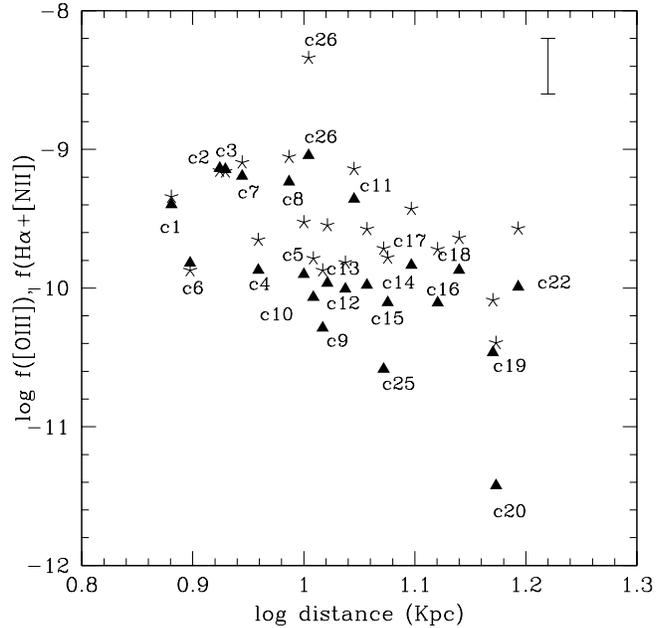}}
\caption{The fluxes of [\ion{O}{III}] (filled triangles) and
[\ion{N}{II}]+\Ha ~(asterisks) observed at the clouds are plotted as function
of the cloud distance from the center.}
\label{f16}
\end{figure}

The emission line flux from the cloud, identified by YOS02 (see their 
Fig. 3) as C20, located at the extreme northern edge of the cone,
is explained by a low flux (log \Fh = 6.5) , \n0= 2 \cm3, and a low shock
velocity (35 \kms).
The largest \Ha + [\ion{N}{II}] flux  is emitted by the cloud identified as C26,
located in the nebula at the  southern edge of the  eastern cloud.
It corresponds to relatively high \Fh~ (log \Fh=9.5) and high shock velocity
($\leq$ 150 \kms).
Also cloud C25 is slightly dislocated from the correlating curve in Fig. 15.
Therefore, it shows  conditions different from those of the other clouds.
Cloud C1 shows the lowest distance from center and is modeled by
log \Fh = 9, \Vs = 70 \kms, and \n0 = 10 \cm3 (Fig.~\ref{f11}).
However, C1 also corresponds to a shock dominated model (Fig.~\ref{f12})
with \Vs=80 \kms and \n0=10 \cm3.
We have not enough data to constrain the model and we conclude that many
different conditions
coexist in the clouds corresponding to the observations; in fact, the flux
from the active nucleus, screened by  the intervening matter, cannot reach
all of the filaments within the observed clouds.
The  geometrical thickness adopted in the models is D= 3 pc,
while the  observed clouds show dimensions of the order of 100 pc.
Shock dominated models are radiation bound, so  the 
results should not change even adopting lower geometrical thickness.

To fit the data
in Figs. 12, 13 and 14, we   multiply the models by factors of 10$^{-5}$,
10$^{-8}$ and 10$^{-6.5}$, respectively. They  account for  the filling factor.
Cloud C20 has a surface
area  of S = 60$\times$ 190 pc$^2$  and the distance from center is r = 14.9
kpc (\cite{Yos02}, Table 2).
From S/(4 $\pi$ r$^2$) $\times$ ff = 10$^{-6.5}$ the   filling
factor results ff = 0.08.
For cloud C26, which corresponds to gas with a higher density, we obtain ff =
1.3 10$^{-4}$.
Regarding cloud C1,  ff= 1.4 10$^{-4}$  results for the radiation dominated
filaments and ff= 0.14 for the shock dominated filaments.

YOS02 claim that the inner region (r$<$ 12 kpc) of
the VEELR  may be excited by nuclear ionizing radiation, while the
excitation mechanism of the outer region (r$>$ 12 kpc) is unclear.
Our analysis is not complete because the [\ion{O}{III}] line fluxes were not
observed for r$>$ 15.6 kpc.
It is found, however,  that clouds C19-C22 correspond to lower fluxes and lower
velocities, because the nuclear flux from the AC is diluted by distance.
The ionization parameter U  that results from modeling the VEELR ranges
between 10$^{-5}$ and 1.5 10$^{-3}$.
The  shock velocities (50-150 \kms) are in agreement with  the radial
velocities in the N-E plume, i.e. 100-200 \kms (\cite{Yos02}).

\section{The  SED of the continuum}

The consistency of the models is checked in Fig. 17 where the SED of the
bremsstrahlung radiation  and  re-radiation by dust  from different clouds
within NGC 4388 is compared with the data.
Therefore, in Fig. 17 two lines correspond to each model, one indicating the
bremsstrahlung from the gas and the other emission from dust in the infrared.
The observational data (from the NED\footnote{NASA/IPAC Extragalactic
Database}) are taken from Boroson \ea (1983), Scoville \ea (1983),
de Vaucouleurs \& Longo (1988),
Soifer \ea (1989), Moshir \ea (1990), Becker \ea (1991),
de Vaucouleurs \ea (1991), Gregory \& Condon (1991), Kinney \ea (1993),
Fabbiano \ea (1992), Zwicky \ea (1961), Douglas \ea (1996),
Dressel \& Condon (1978), Spinoglio \ea (1995) and Mould \ea (1980).
The two data in the soft X-ray range come from Einstein
(Fabbiano \ea 1992), integrated on 0.2-4.0 keV,
and from ROSAT (Halderson \ea 2001), integrated on 0.1-2.4 keV.

\begin{figure}
\resizebox{\hsize}{!}{\includegraphics{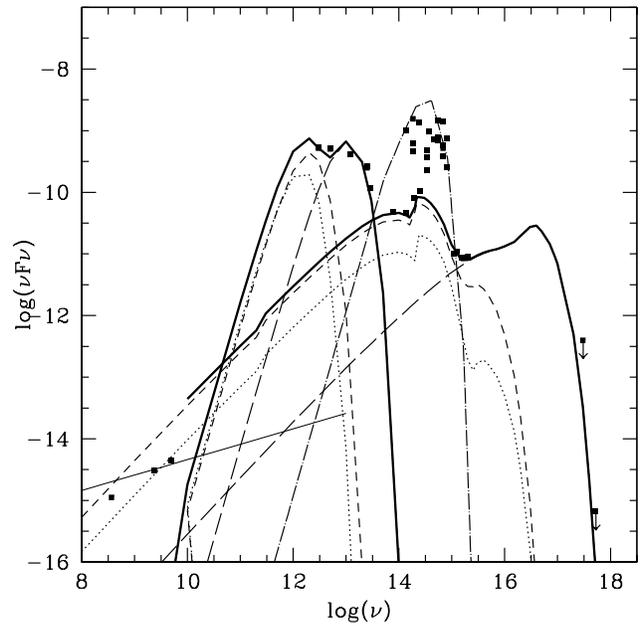}}
\caption{The SED of the continuum.
The data are represented by filled squares.
$\rm V_s=100~km~s^{-1}$, $\rm n_0=100~cm^{-3}$, $\rm log F_H =8.77$,
D=0.01 pc, d/g= 1.5 $10^{-14}$ (short-dash lines),
V$_s$=100 km s$^{-1}$, n$_0$=30 cm$^{-3}$, D=1.3 pc, log \Fh = 8.95
d/g=10$^{-14}$ (dotted lines), and $\rm V_s=500~km~s^{-1}$,
$\rm n_0=300~cm^{-3}$, D=3.3 pc, d/g= 1.5 $10^{-13}$ (long-dash lines).
The summed continuum SED (thick solid lines); radio synchrotron continuum
(thin solid line).}
\label{f17}
\end{figure}

Indeed, the data  from the NED were taken with different precisions and in
different epochs.  Nevertheless, they are very significant because they
 correspond to the  main contributions to the continuum in the different
frequency ranges by
1) bremsstrahlung from the photoionized gas
($\nu$ $\sim$ 5 10$^{13}$- 5 10$^{15}$ Hz),
2) bremsstrahlung from the  hot gas in the downstream region of
shocked clouds ($\nu$ $>$ 10$^{16}$ Hz), 3) emission from the old star population 
($\nu$ $\sim$ 5 10$^{13}$-10$^{15}$ Hz),
4) dust reradiation in the infrared range ($\nu$ $\sim$ 10$^{12}$-10$^{14}$ Hz), 
and 5) radio emission ($\nu$ $<$ 10$^{11}$ Hz).

Besides the line and continuum spectra, the SUMA code calculates
reradiation from dust in a consistent way. Dust is  heated by
the nuclear radiation flux; moreover, gas and dust mutually heat each other by
collisions in shock turbulent regimes (Viegas \& Contini 1994).
Therefore, the frequency corresponding to the dust reradiation peak depends on the
shock velocity. At high \Vs ~the grains will reach high temperatures in the
post shock region and dust reradiation will peak in the mid-IR, while at low
velocities the peak appears in the far-IR.
The peak intensity depends on the dust-to-gas ratio.

Sputtering of the grains, which depends on the shock velocity,
 is calculated throughout the clouds. Small grains are rapidly sputtered,
therefore we adopt silicate grains with an initial radius  of 0.2 $\mu$m.

Emission in the radio range corresponds to
bremsstrahlung from cool gas as well as to synchrotron radiation due to Fermi 
mechanism at the shock front.

The continuum SED corresponding to single clouds are shown in Fig. 17 in order 
to explain the role of the different models. Then, the contributions from all 
the clouds are  summed up to fit the data consistently in all the frequency 
domains.

The models  obtained for the S-W cone  are used to model the SED of NGC 4388,
because the emission in the VEELR is very low and can hardly affect the
continuum observed from the entire galaxy.
The  models presented in Fig.  17 are selected from those which better fit
the bulk of the data shown in  Fig. 7.
We will consider three significant models, regarding the shock velocity,
the preshock density, the geometrical thickness of the cloud, and the
dust-to-gas ratio.
The short-dashed lines refer to a model calculated with \Vs=100 \kms, \n0=100
\kms, D=0.17 pc and d/g=1.5 10$^{-14}$.
Notice that d/g = 10$^{-14}$ by number corresponds to 4.1 10$^{-4}$ by mass
(Draine \& Lee 1984) which is the dust-to-gas ratio in the Galaxy
adopting a silicate density of 3 g \cm3.
Dotted lines represent the SED of the model calculated with D=10 pc,
\n0=30 \cm3, log \Fh = 9.8, and d/g=$10^{-14}$.
Considering that the FWHM of the line profiles show velocities up to
$\geq$ 500 \kms, we have added in Fig. 17 also the continuum of a
SD model corresponding to \Vs = 500 \kms, \n0=300 \cm3, and
d/g =1.5 10$^{-13}$, in order to fit the data  at higher frequencies.
This relatively high shock velocity leads to a temperature in the
immediate postshock region of $\sim$ 3.7 10$^6$ K.
Consequently bremsstrahlung radiation peaks in the soft X-ray domain.

Indeed, the analysis of the ROSAT HRI observations by Matt et al
(1994) has  shown that the 0.1-2.4 keV X-ray emission is 
dominated by an extended component ($\sim$ 4.5 pc). 
It is interpreted as {\it thermal emission from a
collisionally heated plasma and a collection of discrete sources.}
The corresponding  temperature of $\sim$ 0.4 keV
($\sim 4.6 \times 10^6$ K) is in agreement with  the temperature of
$\sim$ 4 10$^6$ K calculated   by \Vs=500 \kms.
No correlation
of the soft X-ray emission with the optical ionization cones
has been found (Matt et al.). This confirms our previous results (Sects. 2 and 3)
showing that rather low ($\sim$ 100 \kms) shock velocities
dominate in the  cones.

Fig.  17 shows that the SED of single models calculated with \Vs=100 \kms~
nicely fit  the data corresponding to the
lowest fluxes in the optical-UV range. The data between 10$^{14}$
and 10$^{15}$ Hz are, however,  contaminated by radiation
from old population stars and are explained by a black body model 
corresponding to 5 10$^3$ K  (dash-dotted line).
This is common to the continua of Seyfert 2 galaxies (\cite{cv00}).

Recall that dust and gas mutually heat each other across the shock front,
therefore, dust re-radiation  from high velocity clouds
(\Vs = 500 \kms) peaks in the mid-IR (see \cite{co02}).
Our results show that these clouds are rather dusty.

The data in  the radio range do not follow exactly the slope corresponding to
bremsstrahlung from a cool gas. Yet, they do not fit a power-law distribution
(thin solid line), which could be explained by synchrotron radiation created
by Fermi mechanism at the shock front.
Fig. 17 shows that both mechanisms contribute to the radio emission flux.

Finally, the thick solid lines in Fig. 17 represent the weighted sum of the
single-cloud models for gas bremsstrahlung and dust reradiation, separately.
The line is interrupted  at frequencies lower than 10$^{10}$ Hz, because
self-absorption, which increases with wavelength, strongly reduces
bremsstrahlung emission in the radio domain.

The model  best fit to the data is obtained adopting
different weights for different models. The  weights  account for
the relative number of clouds corresponding to the model.

The ratio of the weights used to fit the data by the three models is
1 : 0.2 : 0.03, the maximum weight corresponding to \Vs=100 \kms
and D=0.17 pc and the minimum to \Vs=500 \kms.
This confirms that velocities of $\sim$ 100 \kms and relatively
small D ($<$ 1 pc) dominate.

\section{Concluding remarks}

The central regions of the Seyfert-2 galaxy NGC 4388 have been observed by
means of the integral field spectroscopy technique.
The [\ion{O}{III}] \onel5007/\Hb map revealed part of the S-W
ionization cone, with P.A. $\sim$200\degr~ and aperture angle of
$\sim$70\degr. \\
The analysis of the spectra  within the cone shows the
non-thermal active nature of the emission line regions, whose ionization
degree remain high and almost constant even at large distances from the
nucleus.
With simple energy balance calculations we estimate the number of the
ionizing photons produced by the engine and we demonstrate that even if the
photoionization is the main physical process responsible for the excitation
of the gas inside the cone, it cannot be considered the only one under way,
since it cannot explain the observed high ionization far from the source.

We  verify our hypothesis by applying to our spectroscopic data the
SUMA code, which accounts for the combined effect of the photoionization from
an external source and of the shocks.
The results of the models, whose reliability  is cross-checked by comparing
them with the observed SED of the continuum, show that the ionization cone is
characterized by a mixture of low density colliding clouds with large 
(D $\geq$ 1 pc) and small (D$<$0.1 pc) geometrical thickness.
Collisions, in fact, cause turbulence and R-T and K-H instabilities which lead
to fragmentation of matter.
In particular we are able to explain the high [\ion{O}{III}]/\Hb line ratios
observed in the outer edges of the S-W cone by the coupled effect of a low
density gas in  large clouds (D $\geq$ 1 pc).

The models  show that the bulk of the emitting clouds within the S-W cone has
shock velocity of $\sim$ 100 \kms,
definitively lower than those corresponding to the observed FWHM of the
line profiles (200 - 700 \kms).
Adopting the hypothesis  that the extraplanar material in NGC 4388 may
represent the tidal debris  from a recent encounter with a small gas-rich
galaxy (\cite{pog88}; \cite{vei99}), we suggest that \emph{head-on-back} 
shocks accompany the outwards motion of the clouds, which result when an 
incoming gas catches up with preexistent gaseous clouds moving in the same 
direction with a different velocity.

To complete the analysis of the biconical structure of NGC 4388 we have
modeled the  emission line fluxes observed by YOS02 in the VEELR 
of the N-E cone.
We find that the physical conditions are  in agreement with those
found in the S-W cone, taking into consideration dilution with distance
from the AGN.
The results lead to log \Fh = 8.3 - 10.5 (U = 2 10$^{-5}$ - 5 10$^{-4}$),
\n0 = 30 - 100 \cm3, and \Vs $\sim$ 100 -300 \kms~
in the S-W cone, and log \Fh = 6.5 - 9.5 (U = 10$^{-5}$ - 1.5 10$^{-3}$),
\n0 = 2 - 10 \cm3, and \Vs = 30 - 150 \kms~ in the VEELR of the N-E cone.
We prefer to compare the radiation fluxes instead of the ionization parameter, 
because U depends on the density of the gas in the edge of the cloud
facing the radiation source. So, considering that the density decreases
outward the galaxy, dilution is less prominent.
The values of U in the nuclear region,  calculated by Petitjean \& Durret
(1993), are much higher, U =\ 5 10$^{-3}$ - 0.014, indicating that different
conditions characterize the gas in this region.

Concluding, densities, shock velocities, and the magnetic field are lower
in the VEELR, suggesting that beyond the  edges of the VEELR  the emitting clouds
are merging with  the ISM.

\vskip 50pt

\begin{acknowledgements}

We are very grateful to the referee, D. Alloin, for precious comments
which highly improved the presentation of the paper.

M. Contini is grateful to the Astrophysikalisches Institut Potsdam and
the Department of Astronomy Padova for warm hospitality.
We are grateful to Prof. V. Afanasiev and Dr. S. Dodonov of the Special
Astrophysical Observatory (Russia) for their decisive contribution in
observations and useful discussions about data reduction and analysis.

This research has made use of the NASA/IPAC Extragalactic Database (NED) which
is operated by the Jet Propulsion Laboratory, California Institute of
Technology, under contract with the National Aeronautics and Space
Administration.
\end{acknowledgements}

\end{document}